# NbOx based memristor as artificial synapse emulating short term plasticity

Sweety Deswal,[a,b] Ashok Kumar,[a,b] and Ajeet Kumar[a,b*]

[a]Academy of Scientific and Innovative Research (AcSIR), Ghaziabad 201002, India.

[b]CSIR-National Physical Laboratory, Dr. K. S. Krishnan Marg, New Delhi 110012, India

[*]kumarajeet@nplindia.org

## Abstract

Memristors can mimic the functions of biological synapse, where it can simultaneously store the synaptic weight and modulate the transmitted signal. Here, we report $Nb/Nb_2O_5/Pt$ based memristors with bipolar resistive switching, exhibiting synapse like property of gradual and continuously change of conductance with subsequent voltage signals. Mimicking of basic functions of remembering and forgetting processes of biological brain were demonstrated through short term plasticity, spike rate dependent plasticity, paired pulse facilitation and post-titanic potentiation. The device layer interface tuning was shown to affect the device properties shift from digital to analog behaviour. Demonstration of basic synaptic functions in the NbOx based devices makes them suitable for neuromorphic applications.

# Introduction

Conventional digital computing is suffering from the von Neumann bottleneck due to the complex process of data fetch from storage, decode at the process level, and execution by the processors, making them highly energy inefficient when complexity of computation increases.[1-3] This poses one of the major challenges in processing large amount of data. Therefore, new computing architectures inspired by our biological brain are required where the co-location of memory and logic effectively handles the complicated tasks.[4-6] The neural network in our brain, in its simplest form, is a set of neurons connected by weighted synaptic connections.[7] This offers highly parallel processing power to perform various functions like memorization and forgetting based on the modulation of strength or weight of synapse.[8-10] Therefore, to achieve efficient neuromorphic systems, the development of artificial synapse is one of the crucial steps.

Memristors having two terminal structure with a metal-insulator-metal stack can mimic the functions of biological synapse, where it can simultaneously store the synaptic weight and modulate the transmitted signal.[11] The potential of resistive switching devices or memristors to function as a synapse has been demonstrated using various materials like Ag/Si mixture,[12] $Ag_2S$,[13] $GeSb_2Te_2$,[14] HfOx/AlOx bilayer,[15] $WO_x$.[16] The ability of biological synapse to modulate strength of synapse is called synaptic plasticity. Depending on the strength and repeating frequency of input signals, either short term plasticity (STP) or long term plasticity (LTP) can be achieved.[13,17] Spike-timing dependent plasticity (STDP) and spike-rate-dependent plasticity (SRDP) are important synaptic modification rule to be realized by artificial synapses.[18,19] Despite, several materials have been shown to mimic synaptic behaviour, the search for a single material that can have all the desired device properties together is still an important endeavour.

In this work, $Nb_2O_5$ based resistive switching cells ($Nb/Nb_2O_5/Pt$) are investigated to exhibit their potential for artificial synapse applications. Their conductance was found to be changing gradually and continuously with subsequent voltage signals, which is a necessary condition for an electronic device to emulate the functions of biological synapse. We demonstrate the basic synaptic functions including short term plasticity and SRDP with paired-pulse facilitation (PPF) and post-titanic potentiation (PTP) characteristics. We also analysed the correlation of interface conditions with analog behaviour of the device.

**Methods**

A thin niobium pentoxide ($Nb_2O_5$) film ~30 nm was deposited on a $Pt/SiO_2/Si$ substrate using reactive dc magnetron sputtering technique.[20,21] The Platinum (Pt) was masked with a NiCr alloy strip to keep Pt available for probing as bottom electrode. The thin film deposition of $Nb_2O_5$ was performed at 50 W power and at a deposition pressure of $\sim 2.0 \times 10^{-2}$ mbar using a mixture of 94% Ar and 6% $O_2$. The base pressure of the deposition chamber was $\sim 7.0 \times 10^{-7}$ mbar. The as-deposited Nb oxide films were amorphous. Subsequently, an array (5×5) of ~80 nm thick Nb top electrode was deposited with dimension of a×a $\mu m^2$ (a= 100, 200, 300, 400, 500) using shadow mask in ultra high vacuum (UHV) sputtering at base pressure of $\sim 2.0 \times 10^{-8}$ mbar with Ar atmosphere of pressure $6 \times 10^{-3}$ mbar. The electrical dc measurement of $Nb/Nb_2O_5/Pt$ devices was performed at room temperature using two probe arrangement with Keithley 2450 sourcemeasure unit. The bottom electrode was grounded and bias was applied to the top electrode for all dc electrical measurements.

**Results and Discussion**

The $Nb/Nb_2O_5/Pt$ devices exhibited analog bipolar resistive switching. Figure 1(a) shows semi-logarithmic I-V hysteresis loop with positive and negative voltage sweeps. The current in negative voltage sweep region was two orders smaller than that with positive voltage sweep. Such asymmetry in current hysteresis with opposite voltage sweeps is a result of unequal potential barriers at the Nb/$Nb_2O_5$ (0.4 eV) and the Pt/$Nb_2O_5$ (1.75 eV) interfaces. Here, the potential barrier values for discussed interfaces are taken from earlier reports.[22,23] Also, similar asymmetry of I-V curves has been reported by Dong et al.[24] The hysteretic currents for the device operation was in the range of nano to micro ampere, which makes it advantageous in low power (< μW) operation. However, high operating voltage limits their compatibility with current CMOS. No electroformation was required for these analog I-V characteristics in the devices. Figure 1(a) shows that the device exhibited excellent endurance characteristics for at least 100 sweep cycles with ±8 V sweeps with sequence 0V→8V→0V→ -8V→ 0V. In another experiment, where only positive voltages were applied with increasing amplitudes in consecutive sweeps (5 V to 11 V in steps of 1 V), a corresponding increase in hysteretic current was observed Fig. (1b). Area of observed hysteretic loop increased due to increase in current with increase in amplitude of applied voltage. We also studied the current variation with applied voltage pulses of increasing amplitude, shown in Fig 1(c). Positive voltages with increasing amplitudes were applied during ON time successively, while, a read voltage of 1V was applied during OFF time. An increase in current was observed with successive increase in voltage. The current was found to decay to the original value during OFF time of the cycle. This behaviour indicates volatile nature of these devices. We observed a continuous increase in current during every single voltage pulse (ON time). This behaviour is further explored in the manuscript. This volatile, continuous and multilevel change in conductance of device makes it valuable for neuromorphic computing.[25-27]

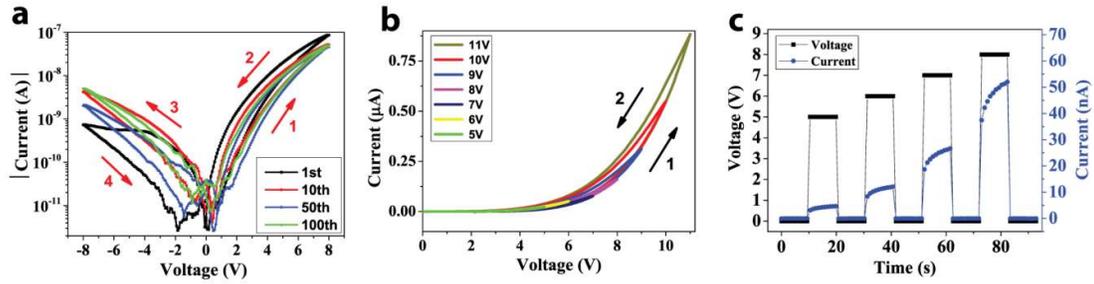

**FIG. 1.** (a) The Fig. shows the reproducible switching for $1^{st}$, $10^{th}$, $50^{th}$ and $100^{th}$ cycles with sweeps of voltage amplitude ±8 V. Here, modulus of current is plotted for –ve voltage sweep. (b) The hysteretic current increases with subsequent voltage sweeps of increasing amplitude from 5 V to 11 V in steps of 1 V. (c) Current increases during ON time with increase in magnitude of voltage. During OFF time, the applied voltage is almost zero (1 V) and current also decays to zero. Curvatures of the traces appear opposite in fig. 1a and 1b due to logarithmic and linear scales of y-axis, respectively.

We also observed an increase in hysteretic current with successive voltage sweeps of same amplitude. The Fig. 2(a) exhibits the gradual increase in conductance of the device with successive positive voltage sweeps of amplitude +14 V. When negative voltage sweeps of same amplitude were applied, the conductance decreased in similar fashion as shown in Fig. 2(b). The variation of current and voltage with respect to the time observed in Fig. 2(a) and (b) is represented as triangular wave form in Fig. 2(c) and (d), respectively. The conductance was gradually increasing (or decreasing) during positive (or negative) voltage sweeps. Due to multilevel and incremental change in conductance, the device seems to be suitable to emulate the functions and plasticity of a biological synapse. Based on the change in the weight of a biological synapse, various functions like learning, forgetting and memorizing are performed by biological synapse. The slope of I-V in subsequent sweeps (Fig. 2(a)) starts from where the last sweep left off. This resembles the process of remembering with continuous repetitions. Here, the current was increasing gradually with continuous repetition of same amplitude voltage sweeps. In similar way, the current was decreasing with subsequent negative voltage sweeps (Fig. 2(b)) resembling the process of forgetting.

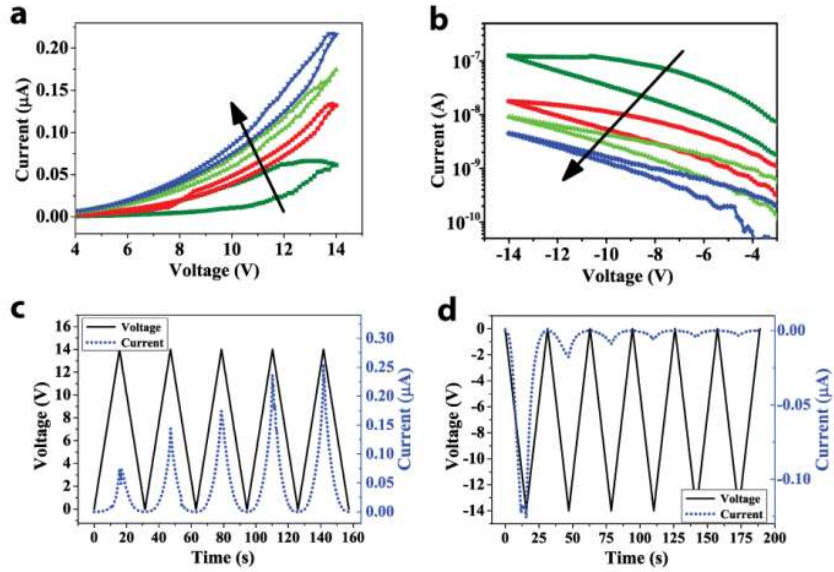

**FIG. 2.** Memorization characteristics of $Nb_2O_5$ based memristor (a) The hysteretic current increases with subsequent voltage sweeps of amplitude 14 V. (b) There is decrease in hysteretic current subsequently with negative voltage sweeps of amplitude -14 V. (c) The plot of hysteretic current and voltage w.r.t time of Fig. (a). (d) shows the current and voltage plots w.r.t time of Fig. (b). There is an increase and decrease in hysteretic current with successive positive and negative voltage sweeps of same amplitudes.

For application as artificial electronic synapse, change in conductance of memristor device was studied with pulse voltage signals. A sequence of 500 identical positive pulses of amplitude 6 V, 8 V and 10 V of width 0.1 s was applied, followed by a sequence of 500 negative similar pulses. A gradual nonlinear increase (decrease) in current was observed with positive (negative) pulses. The current was measured at a read voltage of 3 V after each pulse and the plot of the read current is shown in Fig. 3(a). The current was found to increase and decrease non-linearly with positive and negative pulses, which is called potentiation and depression, respectively. We further investigated this non-linear increase in conductance with respect to different pulse widths, pulse intervals, pulse amplitudes and number of pulses, which is important for a device in order to emulate biological synaptic behaviour. The Fig. 3(b) shows the effect of 8 V voltage pulse with different pulse widths of 0.1 s, 0.5 s and 1 s over the device conductance. We observed that conductance increased with the increase in the pulse width, when the number of pulses was kept fixed. This implies that a particular memory state can be reached with lesser number of pulses having longer pulse widths. This can be related to the fact that if something is learnt continuously for longer time, the probability of remembering it would be higher i.e. a transition towards long term memory.

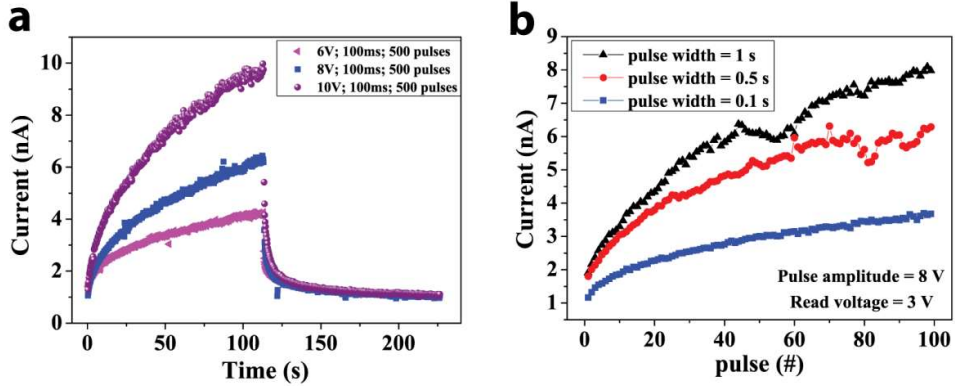

**FIG. 3.** (a) The potentiation and depression characteristics of Nb/Nb$_2$O$_5$/Pt based memristor. A series of 500 positive pulses of width 0.1 s with amplitudes of 6 V, 8 V and 10 V, and subsequently for similar 500 negative pulses was applied to the device. The current gradually increased and then decreased with the application of positive and negative pulses, respectively. The current was measured at a read voltage of 3 V after each pulse and the read current was plotted. (b) The effect of different pulse widths i.e. 0.1 s, 0.5 s and 1 s with the voltage amplitude of 8 V over the current of device for 100 pulses is exhibited. The increase in current was more with larger pulse width.

The interval time between successive pulses also affected the memorization ability of the devices. Fig.s 4(a) – (d) show the change in current when ten identical pulses of amplitude 10 V and width 0.5s with different pulse intervals of 20 s, 10 s, 5 s and 2 s were applied. Fig. 4(e) shows the current obtained with each pulse application for different pulse intervals. With 20 s and 10 s pulse intervals, no increase or insignificant rise in current was observed. The increase in current was significant with 5 s pulse interval. With further reduction in the pulse interval upto 2 s, we observed a very clear and evident increase in current. These results suggest that a memory cannot be achieved at all if the pulses are applied after very large intervals (e.g. 10 s or more for the devices). Further, these measurements were analysed as change in current between first and second pulses, known as paired-pulse facilitation (PPF) and difference between tenth and first pulse currents known as post-titanic potentiation (PTP), shown in Fig. 4(f).

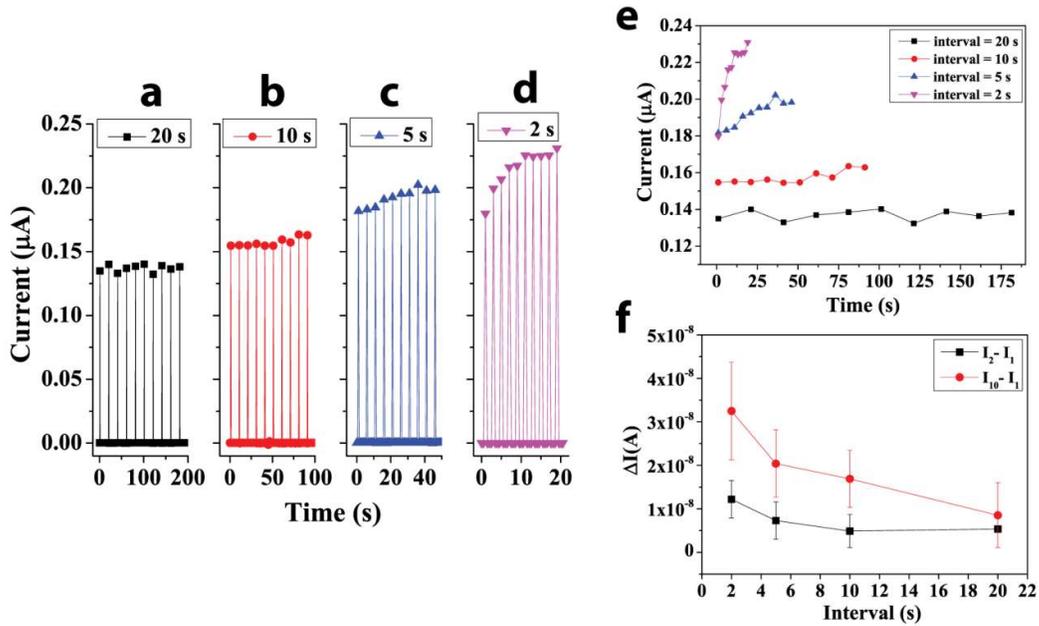

**FIG. 4.** the change in current when 10 identical pulses of 10 V for 0.5 s were applied with different pulse intervals of (a) 20 s, (b) 10 s, (c) 5 s and (d) 2 s. When the interval was 20 s and 10 s, no increase in current was observed with successive pulses. However, when the interval was reduced to 5 s, there is some increase is current. With 2 s interval, the increase in current was clearly visible. (e) Current values obtained at each pulse of 10 V for different intervals shown in (a) – (d). (f) The variation of synaptic current w.r.t pulse interval. ($I_2-I_1$) and ($I_{10}-I_1$) corresponds to paired-pulse facilitation (PPF) and post-titanic potentiation (PTP), respectively.

It can be inferred from Fig. 4(f) that at shortest pulse interval of 2 s, the difference in current increase with number of pulses is much larger as compared to the longer pulse intervals. The currents after $2^{nd}$ and $10^{th}$ pulse remain almost same at the highest pulse interval of 20 s. This indicates that the number of pulses have no significant effect for larger pulse intervals. This behaviour is analogous to memorization of human brain which is affected by frequency of input signals. The long term memory can be achieved with either increasing the frequency of inputs i.e. number of pulses with short intervals or by increasing the amplitude.[13,16,28] However, even with 500 successive pulses with 100 ms interval, LTM could not be achieved and the devices remain volatile in nature. The analog I-V characteristics were observed up to voltage sweep of 16 V and beyond this, permanent breakdown was observed in these devices. This showed that no permanent memory or LTM was achieved even with pulses of higher amplitude. The devices were usually destroyed (further no hysteresis seen) at ≥ 17 V, with a visible bubble formation over the devices. Such kind of observation of device failure with bubble formation is also reported in Hf/HfO$_2$/Pt resistive cells.[29]

To understand the increase in current with successive voltage pulses with pulse interval of 2 s or less, we analysed the decay of current after every pulse of amplitude 12 V and width 100 ms, Fig. 5(a) and (b). The current decay was measured with a read voltage of 3 V. With shorter time intervals, the follow up pulse was applied before the current decayed completely, unlike the longer time intervals where the current goes back to original zero state. Also, the decay relaxation time was analysed by fitting exponential decay function:

$$I = I_0 + Ae^{-t/\tau} \qquad (1)$$

Where, $I_0$ is the steady state current, A is constant and $\tau$ is the relaxation time.
With each successive voltage pulses, the relaxation time increased as shown in Fig. 5(c). This infers that the current takes longer time to decay with successive voltages and thus the volatility of the system decreases. Also this indicates that the history of applied pulses do affect the outcome, somewhat similar to memorization process.

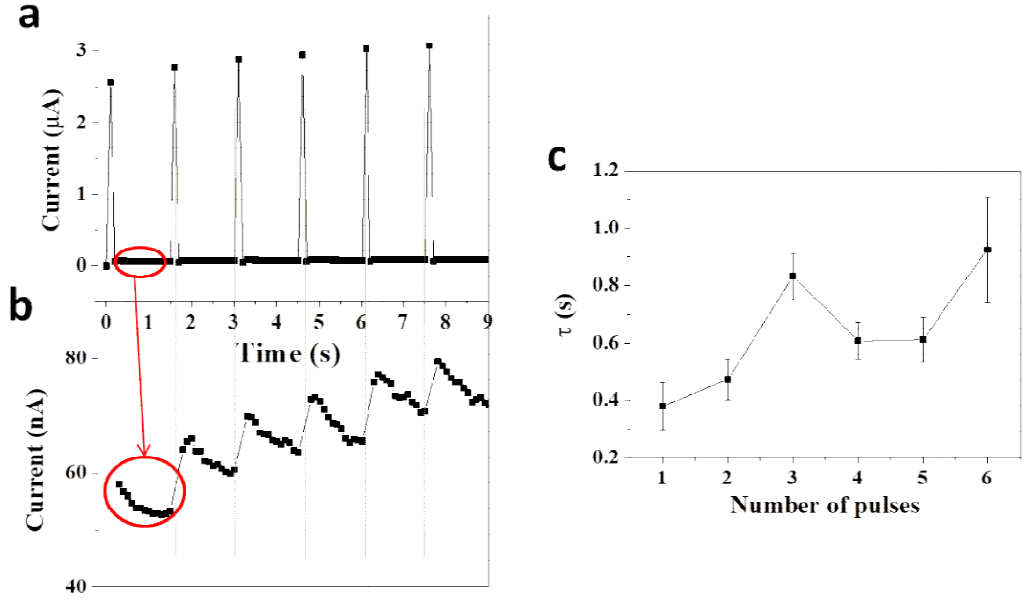

**FIG. 5.** (a) The current increases with subsequent voltage pulses of amplitude 12V and pulse width 100ms applied within an interval of 2s. (b) The decay profile of read current (read at 3 V), which is observed in the intervals between consecutive applied pulses shown in (a). (c) The error bar plot shows that the relaxation time increases with number of pulses, inferring that memorization increases with subsequent pulses.

To investigate the analog switching behaviour of the device, we performed control experiments wherein the top electrode of Nb deposited in ultra high vacuum (2 x $10^{-8}$ mbar) conditions was replaced by $Nb_{low-vac}$ deposited at vacuum of 1 x $10^{-6}$ mbar. Here, Nb and $Nb_{low-vac}$ represent niobium top electrodes deposited at 2 x $10^{-8}$ mbar and 1 x $10^{-6}$ mbar vacuum conditions, respectively. Considering the energy band diagram (Fig. S1), both devices with Nb and $Nb_{low-vac}$ top electrodes were expected to show similar abrupt switching behaviour. However, only $Nb_{low-vac}$/$Nb_2O_5$/Pt devices showed abrupt switching (Fig. S2) similar to our earlier reported devices [30] with Al top electrode, and the devices with Nb showed an analog behaviour with a gradual resistance change. Also, we investigated the conduction mechanism in Nb and $Nb_{low-vac}$ top electrode devices which are shown in Fig. S3. We found that Schottky emission which is an electrode limited conduction was the best fitted mechanism in the devices with Nb top electrode, while the best fitted mechanism for the devices with $Nb_{low-vac}$ top electrode was Poole-Frenkel (which is bulk dominated mechanism). These observations point that the devices with Nb top electrode deposited in ultra high vacuum deposition condition forms a cleaner interface as compared to the $Nb_{low-vac}$ top electrodes, which is responsible for the change in switching behaviour. It is understood that ion migration leads to the formation of a conducting filament in the resistive devices responsible for abrupt switching.[31] Thus, the devices with $Nb_{low-vac}$ top electrode, showing similar abrupt switching, must involve ion migration and filament formation. However, the cleaner interface between Nb and $Nb_2O_5$ show analog behaviour with increasing current, inferring that only electron migration is happening across the interface and the interface barrier, which is high enough to prevent any ion migration and thus, preventing any conducting filament formation.

## Conclusions

The $Nb/Nb_2O_5/Pt$ based memristor showed analog bipolar resistive switching characteristics. The memristor showed capabilities to exhibit various synaptic functions emulating the memorization and forgetting process of biological synapse. We demonstrated short term synaptic plasticity including spike rate dependent plasticity with the effect of pulse width, pulse amplitude, pulse interval and number of pulses. Demonstrating basic synaptic functions along with volatile nature of current-voltage characteristics, the niobium oxide based devices seem suitable for short term synaptic potentiation useful in short term memory and training of neuromorphic computing.

## Supplementary Material

See supplementary material for additional details on abrupt switching, energy band diagram.

## Acknowledgements

S. D. would like to thank University Grant Commission (UGC) under senior research fellowship (SRF) for financial support. This work was financially supported partly by DST INSPIRE Fellowship and partly by CSIR network project AQuaRIUS (PSC0110).